# Least Mean Square/Fourth Algorithm with Application to Sparse Channel Estimation

Guan Gui, Abolfazl Mehbodniya and Fumiyuki Adachi
Department of Communication Engineering
Graduate School of Engineering,
Tohoku University
Sendai, Japan
{gui, mehbod}@mobile.ecei.tohoku.ac.jp, adachi@ecei.tohoku.ac.jp

*Abstract*—Broadband signal transmission over frequency-selective fading channel often requires accurate channel state information at receiver. One of the most attracting adaptive channel estimation methods is least mean square (LMS) algorithm. However, LMS-based method is often degraded by random scaling of input training signal. To improve the estimation performance, in this paper we apply the standard least mean square/fourth (LMS/F) algorithm to adaptive channel estimation (ACE). Since the broadband channel is often described by sparse channel model, such sparsity could be exploited as prior information. First, we propose an adaptive sparse channel estimation (ASCE) method using zero-attracting LMS/F (ZA-LMS/F) algorithm. To exploit the sparsity effectively, an improved channel estimation method is also proposed, using reweighted zero-attracting LMS/F (RZA-LMS/F) algorithm. We explain the reason why sparse LMS/F algorithms using $\ell_1$-norm sparse constraint function can improve the estimation performance by virtual of geometrical interpretation. In addition, for different channel sparsity, we propose a Monte Carlo method to select a regularization parameter for RA-LMS/F and RZA-LMS/F to achieve approximate optimal estimation performance. Finally, simulation results show that the proposed ASCE methods achieve better estimation performance than the conventional one.

## I. INTRODUCTION

Broadband signal transmission is becoming one of the mainstream techniques in the next generation communication systems. Due to the fact that frequency-selective channel fading is unavoidable, accurate channel state information (CSI) is necessary at the receiver for coherent detection [1]. One of effective approaches is adopting adaptive channel estimation (ACE). A typical framework of ACE is shown in Fig. 1. It is well known that ACE using least mean fourth (LMF) algorithm outperforms the least mean square (LMS) algorithm in achieving a better balance between convergence and steady-state performances. Unfortunately, standard LMF algorithm is unstable due to the fact that its stability depends on the following three factors: input signal power, noise power and weight initialization [2].

To fully benefit from the obvious merits of LMS and LMF, it is logical to combine the two algorithms for ACE purposes. The combined LMS/F algorithm was first proposed by Lim and Harris [20], as a method to improve the performance of the LMS adaptive filter without sacrificing the simplicity and stability properties of LMS.

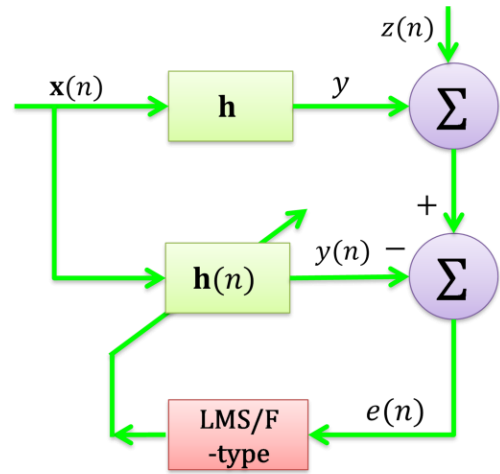

Fig. 1. ASCE for broadband communication systems.

Recently, many channel measurement experiments have verified that broadband channels often exhibit sparse structure. A typical example of sparse system is shown in Fig. 2, where the length of FIR is $N = 16$ while number of dominant coefficients is $K = 2$. In other words, sparse channel is consisted of a very few channel coefficients and most of them are zeros. Unfortunately, ACE using LMS/F algorithm always neglects the inherent sparse structure information and it may degrade the estimation performance. In this paper, we propose sparse LMS/F algorithms with application to ASCE. Inspired by least absolute shrinkage and selection operator (LASSO) algorithm [3], to exploit channel sparsity, $\ell_1$-norm sparse constraint function is utilized in ASCE. Similar to sparse LMS algorithms, two sparse LMS/F algorithms are termed as zero-attracting LMS/F (ZA-LMS/F) and reweighted zero-attracting LMS/F (RZA-LMS/F), respectively.

The main contribution of this paper is proposing the sparse LMS/F algorithms with application to ASCE. Sparse penalized cost functions are constructed for implementing the sparse LMS/F algorithms. Two experiments are demonstrated to confirm the effectiveness of our propose methods. In the first experiment, the average mean square deviation (MSE) performance of sparse LMS/F algorithms is evaluated according to different number of nonzero coefficients. In the second experiment, the MSE performance of propose algorithms is evaluated in different reweighted factors.

The remainder of this paper is organized as follows. A system model is described and standard LMS/F algorithm is introduced in Section II. In section III, sparse ASCE using ZA-LMS/F algorithm is introduced and improved ACSE using RZA-LMS/F algorithm is highlighted. Computer simulations are presented in Section IV in order to evaluate and compare performances of the proposed ASCE methods. Finally, we conclude the paper in Section V.

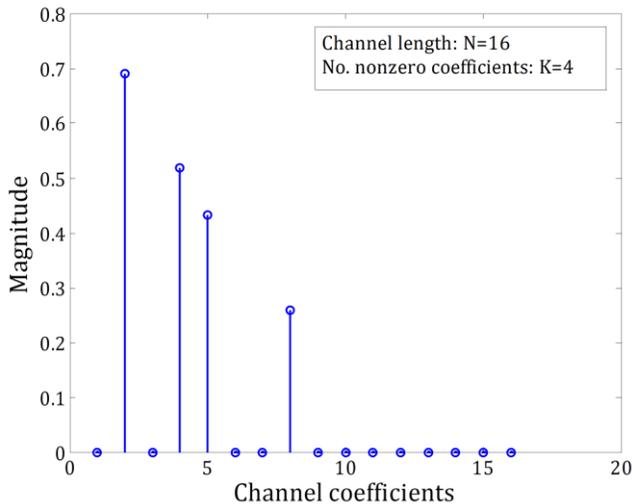
Fig. 2. A typical example of sparse multipath channel.

## II. STANDARD LMS/F ALGORITHM

Consider a baseband frequency-selective fading wireless communication system where FIR sparse channel vector $\mathbf{h} = [h_0, h_1, \ldots, h_{N-1}]^T$ is $N$-length and it is supported only by $K$ nonzero channel taps. Assume that an input training signal $x(n)$ is used to probe the unknown sparse channel. At the receiver side, observed signal $y(n)$ is given by

$$y(n) = \mathbf{h}^T \mathbf{x}(n) + z(n), \quad (1)$$

where $\mathbf{x}(n) = [x(n), x(n-1), \ldots, x(n-N+1)]^T$ denotes the vector of training signal $x(n)$, and $z(n)$ is the additive white Gaussian noise (AWGN) assumed to be independent with $x(n)$. The objective of ASCE is to adaptively estimate the unknown sparse channel estimator $\mathbf{h}(n)$ using the training signal $\mathbf{x}(n)$ and the observed signal $y(n)$. According to [8], we can apply standard LMS/F algorithm to adaptive channel estimation, with the cost function

$$G_{LMSF}(n) = \frac{1}{2}e^2(n) - \frac{1}{2}\lambda \ln(e^2(n) + \lambda), \quad (2)$$

where $\lambda$ is a positive threshold parameter which controls the computational complexity and stability of LMS/F algorithm. With respect to Eq. (2), the corresponding updating equation of LMS/F algorithm is given by

$$\mathbf{h}(n+1) = \mathbf{h}(n) + \mu_f \frac{\partial G_{LMSF}(n)}{\partial \mathbf{h}(n)}$$
$$= \mathbf{h}(n) + \mu_f \frac{e^3(n)\mathbf{x}(n)}{e^2(n) + \lambda}, \quad (3)$$

when $\lambda \gg e^2(n)$, LMS/F algorithm in Eq. (3) behaves like the LMF with a step size of $\mu_f/\lambda$; and when $\lambda \ll e^2(n)$, it reduces to the standard LMS algorithm with a step size of $\mu$. According to the analysis, it is necessary to choose a proper parameter to balance between instability and estimation performance of LMS/F algorithm. Assume the $n$-th received error as $e^2(n) = 0.1$, threshold parameter $\lambda$ controls the variable step-size as shown in Fig. 3. If we fix the $e^2(n)$, then smaller parameter $\rho$ achieves smaller step-size $\mu_f$ which ensures LMS/F more stable and better estimation but at the cost of higher computational complexity, and vice versa.

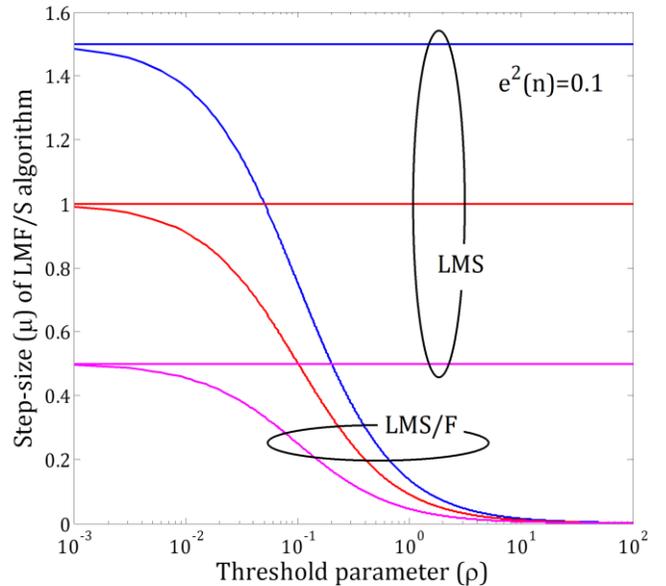
Fig.3. Threshold parameter ($\lambda$) controls the variable step-size of LMS/F algorithm.

## III. SPARSE LMS/F ALGORITHMS

### A. ASCE using ZA-LMS/F algorithm

Recall that the adaptive channel estimation method uses standard LMS/F algorithm in Eq. (2), however, the proposed method does not take advantage of the channel sparsity. This is due to the original cost function in (2) which does not utilize the sparse constraint or penalty function. To exploit the sparsity, we introduce $\ell_1$-norm sparse constraint to the cost function in (2) and obtain a new cost function according to

$$G_{ZA}(n) = \frac{1}{2}e^2(n) - \frac{1}{2}\lambda \ln(e^2(n) + \lambda) + \rho_{ZA}\|\mathbf{h}(n)\|_1, \quad (4)$$

where $\rho_{ZA}$ denotes a regularization parameter which balances the error term, i.e., $1/2e^2(n) - 1/2\lambda \ln(e^2(n) + \lambda)$, and sparsity of $\mathbf{h}(n)$. To better understand the difference between (2) and (4), geometrical interpretation is shown in Fig. 4. Cost function in (2) cannot find sparse solution (convex point) in solution plane. Unlike (2), cost function in (4) can find sparse in solution plane due to its sparse constraint. Hence, the update equation of ZA-LMS/F algorithm is given by

$$\mathbf{h}(n+1) = \mathbf{h}(n) + \mu_f \frac{\partial G_{ZA}(n)}{\partial \mathbf{h}(n)}$$
$$= \mathbf{h}(n) + \mu_f \frac{e^3(n)\mathbf{x}(n)}{e^2(n) + \lambda} + \gamma_{ZA} \text{sgn}(\mathbf{h}(n)), \quad (5)$$

where $\gamma_{ZA} = \mu_f \rho_{ZA}$ and $\text{sgn}(\cdot)$ denotes the sign function which is generated from

$$\text{sgn}(\mathbf{h}(n)) = \frac{\partial \|\mathbf{h}(n)\|_1}{\partial \mathbf{h}(n)} = \begin{cases} 1, & h_i(n) > 0 \\ 0, & h_i(n) = 0, \\ -1, & h_i(n) < 0 \end{cases} \quad (6)$$

where $\mathbf{h}(n) = [h_0(n), \ldots, h_i(n) \ldots, h_{N-1}(n)]^T$ and $i \in \{0,1,\ldots,N\}$. It is well known that ZA-LMS/F uses $\ell_1$-norm constraint to approximate the optimal sparse channel estimation [9].

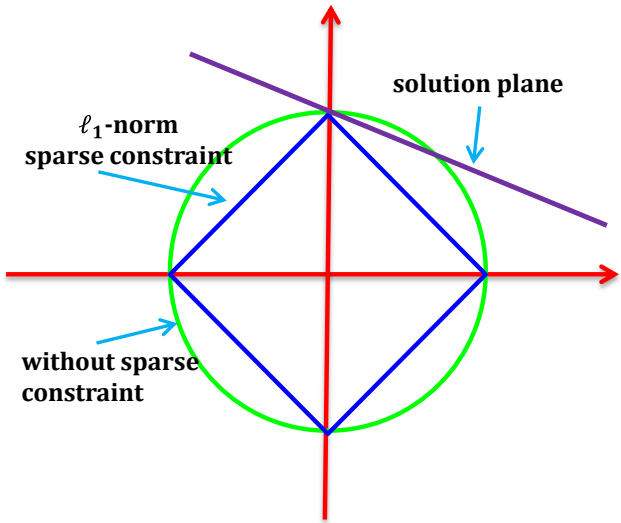

Fig. 4. Sparse channel estimation with $\ell_1$-norm sparse constraint.

### B. Improved ASCE method using RZA-LMS/F algorithm

The ZA-LMS/F cannot distinguish between zero taps and non-zero taps since all the taps are forced to zero uniformly as show in Fig. 5. Unfortunately, ZA-LMS/F based approach will degrade the estimation performance. Motivated by reweighted $\ell_1$-minimization sparse recovery algorithm [10] in CS [11], [12], we proposed an improved ASCE method using RZA-LMS/F algorithm. The cost function of this method is constructed by

$$G_{RZA}(n) = \frac{1}{2}e^2(n) - \frac{1}{2}\lambda \ln(e^2(n) + \lambda)$$
$$+ \rho_{RZA} \sum_{i=0}^{N-1} \log(1 + |h_i|/\varepsilon), \quad (7)$$

where $\rho_{RZA} > 0$ is a regularization parameter which trades off the estimation error and channel sparsity. The corresponding update equation is

$$\mathbf{h}(n+1) = \mathbf{h}(n) + \mu_f \frac{\partial G_{RZA}(n)}{\partial \mathbf{h}(n)}$$
$$= \mathbf{h}(n) + \mu_f \frac{e^3(n)\mathbf{x}(n)}{e^2(n) + \lambda} + \gamma_{RZA} \frac{\text{sgn}(\mathbf{h}(n))}{1 + \varepsilon|\mathbf{h}(n)|}, \quad (8)$$

where $\gamma_{RZA} = \mu_f \rho_{RZA}/\varepsilon$ is a parameter which depends on step-size $\mu_f$, regularization parameter $\rho_{RZA}$ and threshold $\varepsilon$. In the second term of (11), smaller than $1/\varepsilon$ channel coefficients $|h_i(n)|, i = 0,1,\ldots,N-1$ are replaced by zeros in high probability.

### C. Regularization parameter selection for sparse LMS/F algorithms

It is well known that regularization parameter is very important for LASSO based sparse channel estimation [13]. In [14], a parameter selection methods was proposed for LASSO based partial sparse channel estimation. To the best of our knowledge, however, there is no paper report on regularization parameter selection method for ASCE. Here, we propose an approximate optimal selection method by Monte Carlo simulation which adopts 1000 runs for achieving average performance. Parameters for computer simulation are given in Tab. I. The estimation performance is evaluated by average mean square error (MSE) which is defined by

$$\text{Avergae MSE}\{\mathbf{h}(n)\} = \text{E}\{\|\mathbf{h} - \mathbf{h}(n)\|_2^2\}, \quad (9)$$

where $\text{E}\{\cdot\}$ denotes the expectation operator, $\mathbf{h}$ and $\mathbf{h}(n)$ are the actual channel vector and its $n$-th iterative adaptive channel estimator, respectively.

TAB. I. SIMULATION PARAMETERS.

| parameters | values |
|---|---|
| channel length | $N = 16$ |
| no. of nonzero coefficients | $K = 2$ and $4$ |
| step-size | $\mu_f = 0.05$ |
| threshold | $\rho = 0.8$ |
| re-weighted factor for RZA-LMS/F | $\varepsilon = 20$ |

Utilizing different regularization parameters, performance curves of ZA-LMS/F and RZA-LMS/F are depicted in Fig. 5 and Fig. 6, respectively. In Fig. 5, it is easy to find that MSE performance is near optimal when regularization parameters are selected as $\rho_{ZA} = 0.0004$ and $\rho_{ZA} = 0.0002$ for $K = 2$ and $K = 4$, respectively. Likewise, in Fig. 6, choosing approximate optimal regularization parameters $\rho_{RZA} = 0.02$ and $\rho_{RZA} = 0.01$ for RZA-LMS/F can achieve near optimal estimation performance when $K = 2$ and $K = 4$, respectively. Hence, these parameters will be utilized for performance comparison with sparse LMS algorithms.

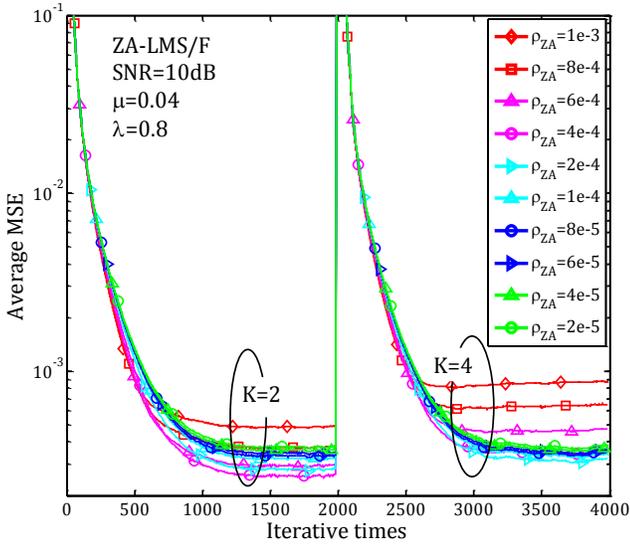

Fig.5. ZA-LMS/F based sparse channel estimation performance depends on regularization parameter $\rho_{ZA}$.

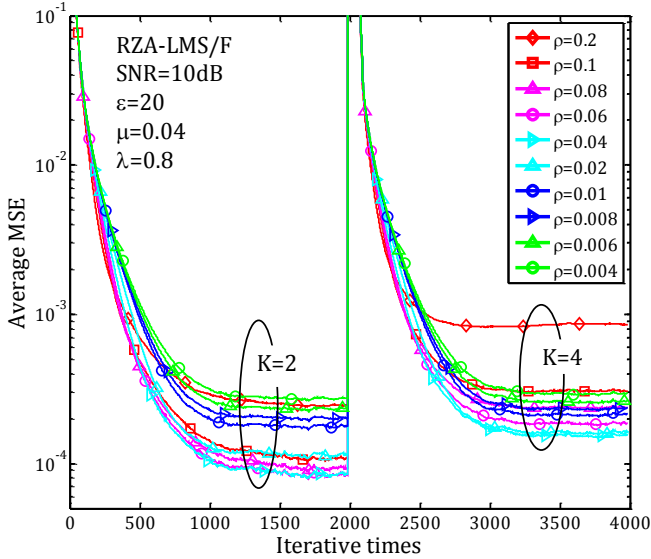

Fig.6. RZA-LMS/F based sparse channel estimation performance depends on regularization parameter $\rho_{RZA}$.

## IV. COMPUTER SIMULATIONS

In this section, the proposed ASCE methods using (R)ZA-LMS/F algorithm is evaluated. To obtain the average performance, 1000 independent Monte-Carlo runs are adopted. The length of channel vector **h** is set as $N = 16$ and its number of dominant taps is set to $K = 2$ and 4, respectively. Each dominant channel tap follows random Gaussian distribution as $\mathcal{CN}(0, \sigma_h^2)$ and their positions are randomly allocated within the length of **h** which is subject to $E\{||\mathbf{h}||_2^2 = 1\}$. The received signal-to-noise ratio (SNR) is defined as $10\log(E_0/\sigma_n^2)$, where $E_0 = 1$ is the unit transmission power. Here, we set the SNR as 10dB in computer simulation. All of the step sizes and regularization parameters are listed in Tab. II.

TAB. II. SIMULATION PARAMETERS.

| parameters | values |
| --- | --- |
| channel length | $N = 16$ |
| no. of nonzero coefficients | $K = 2$ and 4 |
| distribution of nonzero coefficient | random Gaussian $\mathcal{CN}(0,1)$ |
| threshold parameter for LMS/F-type | $\lambda = 0.8$ |
| SNR | 10dB |
| step-size | $\mu_s = \mu_f = 0.04$ |
| regularization parameter for $K = 2$ | $\rho_{ZA} = 0.0004$ and $\rho_{RZA} = 0.04$ $\rho_{ZAS} = 0.008$ and $\rho_{RZAS} = 0.8$ |
| regularization parameter for $K = 4$ | $\rho_{ZA} = 0.0002$ and $\rho_{RZA} = 0.02$ $\rho_{ZAS} = 0.004$ and $\rho_{RZAS} = 0.4$ |
| re-weighted factor for RZA-LMS(/F) | $\varepsilon = 20$ |

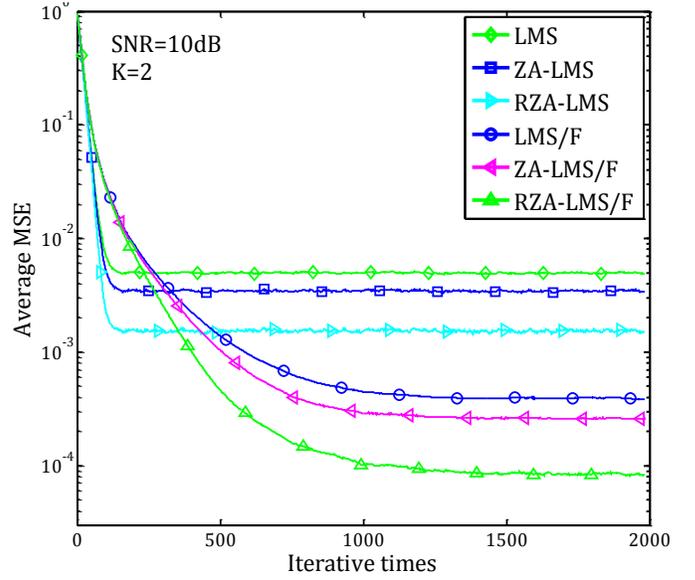

Fig. 7. Performance comparison at $K = 2$.

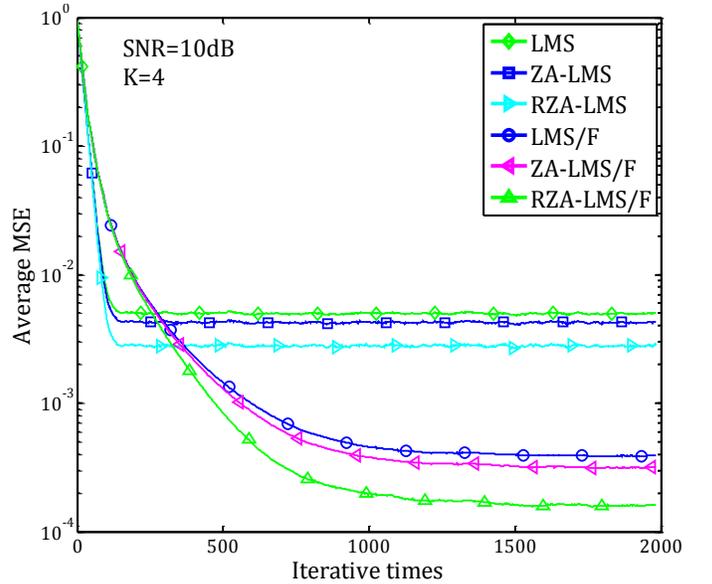

Fig. 8. Performance comparison at $K = 4$.

In the first experiment, average MSE performance of proposed methods is evaluated for $K = 2$ and $4$. To confirm the effectiveness of the proposed methods, we compare them with sparse LMS algorithms, i.e., ZA-LMS and RZA-LMS [15]. For a fair comparison of our proposed methods with sparse LMS methods, we utilize the same step-size, i.e., $\mu_s = \mu_f$. In addition, to achieve approximate optimal sparse estimation performance, regularization parameters for two sparse LMS algorithms are adopted from the paper [16], i.e., $\rho_{ZAS} = 0.008$ and $\rho_{RZAS} = 0.8$ for $K = 2$; $\rho_{ZAS} = 0.004$ and $\rho_{RZAS} = 0.4$ for $K = 4$. Average MSE performance comparison curves are depicted in Fig. 7 and Fig. 8, respectively. Obviously, LMS/F-type methods achieves better estimation performance than LMS-type ones in [15]. According to the two figures, sparse LMS/F algorithms, i.e., ZA-LMS/F and RZA-LMS/F, achieve better estimation performance than LMS/F due to the fact that sparse LMS/F algorithms utilize $\ell_1$-norm sparse constraint function.

In the second experiment, as shown in Fig. 9, estimation performance curves of RZA-LMS/F, utilizing different reweighted factors $\varepsilon \in \{1, 2, 5, \ldots, 50\}$ are depicted for $K = 2$ and $4$. If we set the numerical values for parameters similar to Tab. II, when $K = 2$, RZA-LMS/F using $\varepsilon = 20$ or $25$ can achieve approximate optimal estimation performance. Fig. 9 shows that RZA-LMS/F algorithm depends on reweighted factor. Hence, the proper selection of the reweighted factor is also important when applying the RZA-LMS/F algorithm in adaptive sparse channel estimation.

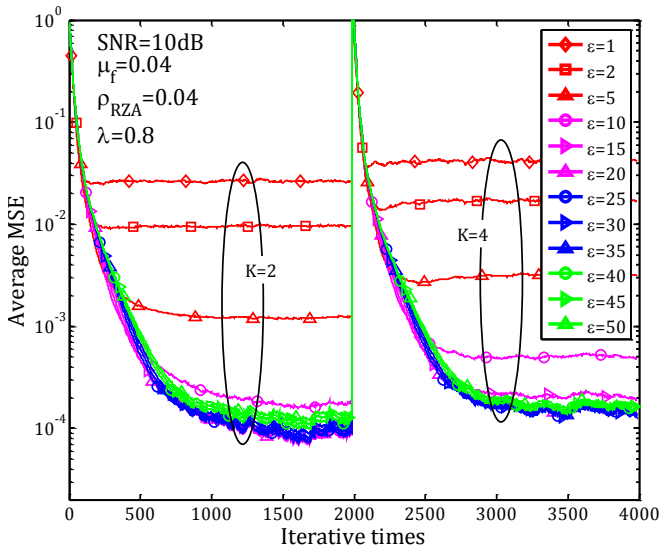

Fig. 9. RZA-LMS/F based sparse channel estimation performance depends on re-weighted factor ε.

## V. CONCLSION

In this paper, a novel LMS/F algorithm was applied in ASCE. Based on the CS theory, we first proposed a novel ASCE method using ZA-LMS/F algorithm. Inspired by re-weighted $\ell_1$-norm algorithm in CS, an improved ASCE method using RZA-LMS/F algorithm was proposed. By Monte Carlo simulation, we proposed a simple method for choosing the approximate optimal regularization parameter for sparse LMS/F algorithm, i.e., ZA-LMS/F and RZA-LMS/F. Simulation results showed that the proposed ASCE methods using ZA-LMS/F and RZA-LMS/F algorithms achieve better performance than any sparse LMS methods.

ACKNOWLEDGMENT

This work was supported in part by the Japan Society for the Promotion of Science (JSPS) postdoctoral fellowship and the National Natural Science Foundation of China under Grant 61261048.